\documentclass[aps,pre,groupedaddress,preprint,longbibliography]{revtex4-1}
\usepackage{graphicx}
\usepackage{epstopdf, epsfig}
\usepackage{dcolumn}
\usepackage{bm}
\usepackage{xcolor}
\usepackage{algpseudocode} 
\usepackage[]{algorithmicx}
\usepackage{varwidth}
\begin{document}
\preprint{}
\title{ Wet to dry self-transitions in dense emulsions: from order to disorder and back}

\author{Andrea Montessori}
\email[]{a.montessori@iac.cnr.it}
\affiliation{Istituto per le Applicazioni del Calcolo CNR, via dei Taurini 19, 00185, Rome, Italy}

\author{Marco Lauricella}
\affiliation{Istituto per le Applicazioni del Calcolo CNR, via dei Taurini 19, 00185, Rome, Italy}

\author{Adriano Tiribocchi}
\affiliation{Center for Life Nanoscience at la Sapienza, Istituto Italiano di Tecnologia, viale Regina Elena 295, 00161, Rome, Italy }
\affiliation{\textcolor{black}{Istituto per le Applicazioni del Calcolo CNR, via dei Taurini 19, 00185, Rome, Italy}}

\author{Fabio Bonaccorso}
\affiliation{Center for Life Nanoscience at la Sapienza, Istituto Italiano di Tecnologia, viale Regina Elena 295, 00161, Rome, Italy }
\affiliation{\textcolor{black}{Istituto per le Applicazioni del Calcolo CNR, via dei Taurini 19, 00185, Rome, Italy}}
\affiliation{\textcolor{black}{Università degi studi Roma "Tor Vergata", Via Cracovia, 50, 00133, Rome, Italy}}

\author{Sauro Succi}
\affiliation{Center for Life Nanoscience at la Sapienza, Istituto Italiano di Tecnologia, viale Regina Elena 295, 00161, Rome, Italy}
\affiliation{\textcolor{black}{Istituto per le Applicazioni del Calcolo CNR, via dei Taurini 19, 00185, Rome, Italy}}
\affiliation{Institute for Applied Computational Science, Harvard John A. Paulson School of Engineering and Applied Sciences, Cambridge,
MA 02138, United States}
\date{\today}

\begin{abstract}

One of the most distinctive hallmarks of many-body systems far from equilibrium is the 
spontaneous emergence of order under conditions where disorder would be plausibly expected.
Here, we report on the self-transition between ordered and disordered emulsions in divergent microfluidic channels,  i.e. from monodisperse assemblies to heterogeneous polydisperse foam-like structures, and back again to ordered ones. 
The transition is driven by the nonlinear competition between  viscous dissipation and surface tension forces as controlled
by the device geometry, particularly the aperture angle of the divergent microfluidic channel. 
An unexpected route back to order is observed in the regime of large
opening angles, where a trend towards increasing disorder would be intuitively expected.

\end{abstract}

\maketitle

\section{Introduction}

Self-organization can be broadly defined as the complex of processes which drives  
the emergence of spontaneous order in a given system, due to the action of local interactions between its 
elementary constituents \cite{prigogine2018order}.  
This concept has provided a major paradigm to gain a deeper insight into a number of phenomena
across a broad variety of complex systems in physics, engineering, biology  and society \cite{rosen2009dendron,camazine2003self,kerner1998experimental,lee2010dynamic,ballerini2008interaction}.  
Self-organization is usually triggered and sustained by competing processes far from equilibrium, as they occur in a gamut of different scientific endeavors, from natural sciences and biology to 
economics and anthropology \cite{karsenti2008self,stanley2002self,wolfram1984cellular,wolfram2018cellular,chopard1998cellular,goldenfeld1999simple}. 
and often efficiently exploited to find innovative design solutions in 
a number of engineering applications \cite{vogel2015color,dinsmore2002colloidosomes,guzowski2015droplet}.
From this standpoint, droplet-based microfluidics, namely the science of generating and manipulating large 
quantities of micron-sized droplets, offers a literal Pandora's box of possibilities to investigate 
the physics of many-body systems out of equilibrium. In particular, the self-assembly between droplets and 
bubbles which results from the subtle multiscale competition between different 
forces and interactions, such as the external drive, interfacial (attractive) 
forces, near-contact (repulsive) interactions, viscous dissipation and inertia \cite{montessori2019mesoscale,montessori2019modeling}.
Most importantly, in many instances, such competition is highly sensitive to 
geometrical factors, primarily the presence of confining boundaries.   
Among others, the ability to manipulate and control tiny volumes of droplets allows the generation 
of highly ordered porous matrices with finely tunable structural parameters \cite{costantini2014}. 
This opens up the possibility of designing  novel families  
of materials with potential use in a wide range of advanced applications, such as catalyst 
supports, ion-exchange modules, separation media and scaffolds in tissue engineering \cite{lee2005scaffold,hollister2005porous,montemore2017effect,montessoriprf}.

Recently, Gai et al.\cite{gai2016spatiotemporal}, reported an unexpected ordering in the flow of a
quasi-2D concentrated emulsion in a convergent microfluidic channel, and showed 
that  confinement of the 2D soft crystal in the extrusion flow causes the reorganization of the crystal internal structure in a highly ordered pattern \cite{gai2016spatiotemporal,gai2019timescale}. 
The self-reorganization of the crystal is expected to bear major implications for the realization of confined low-dimensional 
materials, crucial for applications ranging from optoelectronics to energy conversion, which might be easier to control than previously thought, thus leading to novel flow control 
and mixing strategies in droplet microfluidics.

In this paper, we report on the self-transition between \textit{wet} and \textit{dry} emulsions 
\cite{marmottantprl,garstecki2006flowing}, namely from 
ordered monodisperse assemblies to heterogeneous and polydisperse foam-like structures, in 
divergent microfluidic channel. 

\textcolor{black}{Following the common terminology \cite{marmottantsoft,furuta2016close}, foams are \textit{wet} when their droplets appears nearly round and the structures they form are organized according to ordered hexagonal patterns which flow basically deformation-free. In the \textit{dry} regime instead, the droplets come closer and deform, assuming typical polyhedral shapes  and giving rise to typical disordered foam-like structures.}
\textcolor{black}{Such transition is driven by the capillary number, i.e. the competition between viscous dissipation and
surface tension, which is in turn modulated by the device geometry}.
In particular, we observe a return to an ordered state 
in a parameter regime where a transition towards disorder would be intuitively expected.  


\section{Method}
\textcolor{black}{
In this section, we briefly describe the numerical model employed, namely an extended color-gradient lattice Boltzmann approach with repulsive near-contact interactions, previously introduced in \cite{montessori2019jfm}. In the multicomponent LB model, two sets of distribution functions evolve, according to the usual streaming-collision algorithm (see \cite{succi2018lattice,kruger2017lattice}), to  track the evolution of the two fluid components:
\begin{equation} \label{CGLBE}
f_{i}^{k} \left(\vec{x}+\vec{c}_{i}\Delta t,\,t+\Delta t\right) =f_{i}^{k}\left(\vec{x},\,t\right)+\Omega_{i}^{k}[ f_{i}^{k}\left(\vec{x},\,t\right)] +S_i(\vec{x},t),
\end{equation}
where $f_{i}^{k}$ is the discrete distribution function, representing the probability of finding a particle of the $k-th$ component at position $\vec{x}$, time $t$ with discrete velocity $\vec{c}_{i}$, and $S_i$ is a source term coding for the effect of external forces (such as gravity, near-contact interactions, etc). 
In equation \ref{CGLBE} the time step is taken equal to $1$, and the index $i$ spans over the discrete lattice
directions $i = 1,...,b$, being $b=9$ for a two  dimensional nine speed lattice (D2Q9).
The density $\rho^{k}$ of the $k-th$ component and the total linear momentum of the mixture 
$\rho \vec{u}=\sum_k\rho^{k}\vec{u^k} $  are obtained, respectively, via the zeroth and the first order moment 
of the lattice distributions
$\rho^{k}\left(\vec{x},\,t\right) = \sum_i f_{i}^{k}\left(\vec{x},\,t\right)$ and 
$\rho \vec{u} = \sum_i  \sum_k f_{i}^{k}\left(\vec{x},\,t\right) \vec{c}_{i}
$.
The collision operator splits into three components \cite{gunstensen1991lattice,leclaire2012numerical,leclaire2017generalized}: 
\begin{equation}
\Omega_{i}^{k} = \left(\Omega_{i}^{k}\right)^{(3)}\left[\left(\Omega_{i}^{k}\right)^{(1)}+\left(\Omega_{i}^{k}\right)^{(2)}\right].
\end{equation}
where $\left(\Omega_{i}^{k}\right)^{(1)}$, stands for the standard collisional relaxation \cite{succi2018lattice}, 
$\left(\Omega_{i}^{k}\right)^{(2)}$ code for the perturbation step \cite{gunstensen1991lattice}, contributing to the buildup of the interfacial tension while $\left(\Omega_{i}^{k}\right)^{(3)}$ is the recoloring step \cite{gunstensen1991lattice,latva2005diffusion}, which promotes the segregation between the two species, minimising their mutual diffusion.
A Chapman-Enskog multiscale expansion can be employed to show that the hydrodynamic limit of Eq.\ref{CGLBE} is a
set of equations for the conservation of mass and linear momentum (i.e. the Navier-Stokes equations), with a capillary stress tensor of the form:
\begin{equation}\label{capstress}
\bm{\Sigma}=-\tau\sum_i \sum_k\left(\Omega_{i}^{k}\right)^{(2)} \vec{c}_i \vec{c_i}= \frac{\sigma}{2 |\nabla \rho|}(|\nabla \rho|^2\mathbf{I} - \nabla \rho \otimes \nabla \rho)
\end{equation}
being $\tau$ the collision relaxation time, controlling the kinematic viscosity via the relation  
$\nu=c_s^2(\tau-1/2)$ ( $c_s=1/\sqrt{3}$ the sound speed of the model) and $\sigma$ is the surface 
tension \cite{succi2018lattice,kruger2017lattice}. 
In eq. \ref{capstress}, the symbol $\otimes$ denotes a dyadic tensor product. 
The stress-jump condition across a fluid interface is then augmented with an intra-component repulsive term aimed at 
condensing the effect of all the repulsive near-contact forces (i.e., Van der Waals, electrostatic, steric 
and hydration repulsion) acting on much smaller scales ($\sim  O(1 \; nm)$)  than those resolved 
on the lattice (typically well above hundreds of nanometers).
It takes the following form:
\begin{equation}
\mathbf{T}^1\cdot \vec{n} - \mathbf{T}^2 \cdot \vec{n}=-\nabla(\sigma \mathbf{I} - \sigma (\vec{n}\otimes \vec{n})) - \pi \vec{n}
\end{equation}
where $\pi[h(\vec{x})]$ is responsible for the repulsion between neighboring fluid interfaces, 
$h(\vec{x})$ being the distance between interacting interfaces along the normal $\vec{n}$.\\
The additional, repulsive term can be readily included within the LB framework, by adding a 
forcing term acting only on the fluid interfaces in near-contact, namely:
\begin{equation}
\vec{F}_{rep}= \nabla \pi := - A_{h}[h(\vec{x})]\vec{n} \delta_I
\end{equation}
In the above, $A_h[ h(\vec{x})]$ is the parameter controlling the strength (force per unit volume)
of the near-contact interactions, $h(\vec{x})$ is the distance between the interfaces, $\vec{n}$ is a unit vector normal to the interface and $\delta_I\propto \nabla\phi$ is a function, proportional to the phase field $\phi=\frac{\rho^1-\rho^2}{\rho^1+\rho^2}$, employed to localize the force at the interface.
The addition of the repulsive force (added to the right hand side of Eq.~\ref{CGLBE}) naturally leads to the following (extended) 
conservation law for the momentum equation:
\begin{equation} \label{NSEmod}
\frac{\partial \rho \vec{u}}{\partial t} + \nabla \cdot {\rho \vec{u}\vec{u}}=-\nabla p + \nabla \cdot [\rho \nu (\nabla \vec{u} + \nabla \vec{u}^T)] + \nabla \cdot (\bm{\Sigma}  +   \pi \mathbf{I})
\end{equation}
namely  the Navier-Stokes equation for a multicomponent system, augmented with a surface-localized repulsive 
term, expressed through the gradient of the potential function $ \pi$.}

\section{Results and Discussion}

\begin{figure}
\centering
\includegraphics[scale=1.2]{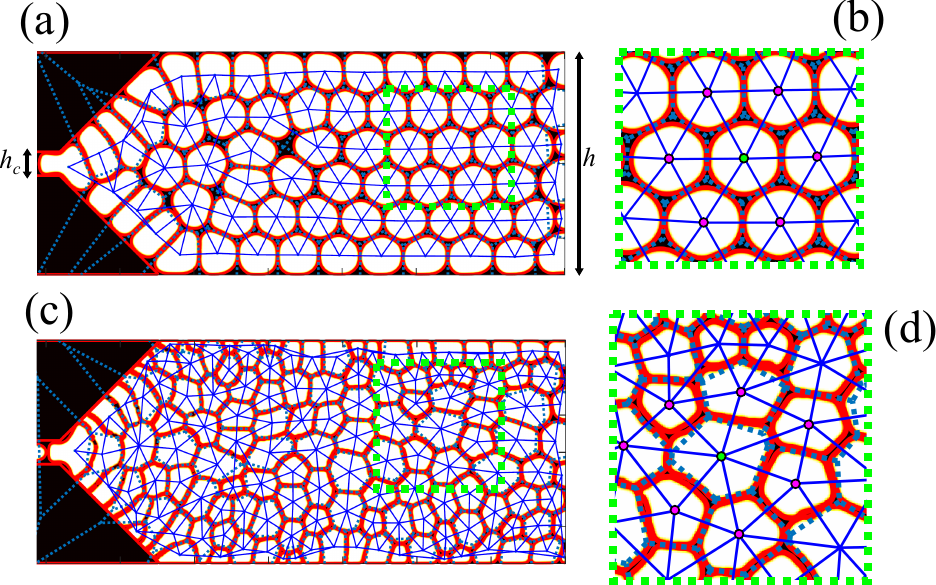}
\caption{Droplet assemblies within a microfluidic channel with a divergent opening angle $\alpha=45^\circ$ for two different Inlet channel Capillary number (a) $Ca=0.04$ (c) $Ca=0.16$. Panel (b) clearly shows the ordered, hexagonal packing typical of wet-state emulsions, while (d) the foam-like structure which results in a neat distortion of the Delauney triangulation (blue solid lines connecting the centers of neighboring  droplets) 
and the associated Voronoi  tesselation (dotted polygons enclosing the droplets) as well.
The red lines are isocontour lines drawn for $\phi_{min}<\phi<\phi_{max}$  being $\phi$ the local phase field $\phi=(\rho_1 -\rho_2)/(\rho_1 + \rho_2)$ ($\rho_i$ the density of the $i$-th phase).
The lines are superimposed to a density field. 
The thickness of the red isocontour line has been widening in order to better visualized the droplet contours.
}
\label{fig0}
\end{figure}

The simulation set-up (see figure \ref{fig0}) consists of a microfluidic device composed by an inlet channel ($h_c$), a divergent channel 
with opening angle $\alpha$ and a main channel connected with the  divergent ($h=10h_c$).
The droplets are continuously generated in a buffer channel, placed upstream the inlet channel whose height is 
$h_{in}\sim1.7h_c$, equal to the droplets diameter. This way, the droplets are forced to deform as they enter the narrower inlet channel, taking a typical oblate shape. 

The fluid motion is driven by a body force which mimics the effect of a pressure gradient 
across the device, which is set in such a way as to guarantee laminar flow conditions within 
both inlet and main channels.

\textcolor{black}{The main parameters employed (expressed in simulation (lattice) units) are the following.\\
The  microfluidic device is composed by  a thin inlet channel (height $h_c=30$, length $l_c=220$) within which droplets are produced and a main, or  self-assembly channel, height $h=10 h_c$, length $l=900$, where the droplets are transported downstream by the main flow and self-assemble in clusters during their motion.\\
The droplet diameter is set to $ D=50 $ lattice units, more than sufficient to capture the complex interfacial phenomena occurring in droplet microfluidics (50 lattice points per diameter means a Cahn number of the order $0.08$, typical in resolved diffuse interface simulations of complex interfaces (see \cite{magaletti2013sharp})).
The motion of the droplet is realized by imposing a constant body force $g=10^{-5}$.
The  viscosity of the two fluids has been set  to  $\nu=0.167$ while the near-contact force has been set to $A_h=0.1$.
The choice of magnitude of the body force, along with the kinematic viscosity of the fluids is such to determine a droplet Reynolds number within the inlet channel $Re\sim2.5$, small enough to guarantee laminar flow conditions.
The surface tension has been varied in the range $\sigma=0.007 \div 0.02$.\\
Finally, the droplet generation is performed by implementing an \textit{internal} periodic boundary condition whose short explanation is reported in appendix.
}

\textcolor{black}{All the simulations were performed in two-dimensions, being this a reasonable approximation for the simulation of droplets' phenomena in shallow microfluidic channels.
We wish to point out that the only parameter which has been varied throughout the simulation is the surface tension between the two components which, in turn, allowed us to tune the capillary number. The rate of injection of both the dispersed and the continuous phase were kept constant in all the simulations.}

In figure \ref{fig0}, we report two different assemblies of droplets within a microfluidic 
channel with a divergent opening angle $\alpha=45^\circ$. 

This figure shows that the tuning of the inlet Capillary number,
($Ca= U_d \mu_d/\sigma$, the $d$ subscript standing for droplet, $U_d$ is the average droplet velocity 
within the inlet channel, $\mu_d$ the dynamic viscosity and $\sigma$ the surface tension of the mixture), allows to switch between a closely 
packed, ordered, monodisperse  emulsion (\ref{fig0}(a)) characterized by regular hexagonal assemblies 
of droplets, traveling along the micro-channel,  to a  foam-like structures, formed by polyhedral-shaped droplets 
(see fig. \ref{fig0} (c)  (d)). 

The resulting structures appear to be irregular and polydispersed, as indicated by the distortion  of the Delaunay triangulation and its dual Voronoi tessellation \cite{lulli2018metastability}.  
\textcolor{black}{We wish to highlight that, both the dispersed and continuous phases' discharges are kept constant in all the simulations. Thus, the observed transition is likely to by due to (i) the breakup processes promoting the formation of liquid films and (ii) the increased deformability of the droplets interface, due to the lower values of surface tension employed. Typically, droplet breakups increase the amount of interface, leading to an augment of the total length of the thin-film and to a redistribution of the dispersed phase in the system. Such dynamics is controlled by the Capillary number whose increase (within a quite wide range of aperture angles) leads to a spontaneous transition from an ordered state to a disordered one displaying the typical features of a dense foam, namely (i) polydispersity, (ii) formation of an interconnected web of plateaus, (iii) departure of the droplets shapes from the circular or spherical one and (iv) formation of droplets assemblies which are not regular as in the wet case. }

The transition between different droplets' structures depends not only
on the inlet capillary number but also on the geometrical details of the device, the latter 
being responsible for a counter-intuitive behavior, to be detailed shortly.  

We begin with a phenomenological description of the droplets injection within the 
diverging channel (for $\alpha=45^\circ$), as influenced by the Capillary number.

\begin{figure}
\centering
\includegraphics[scale=0.6]{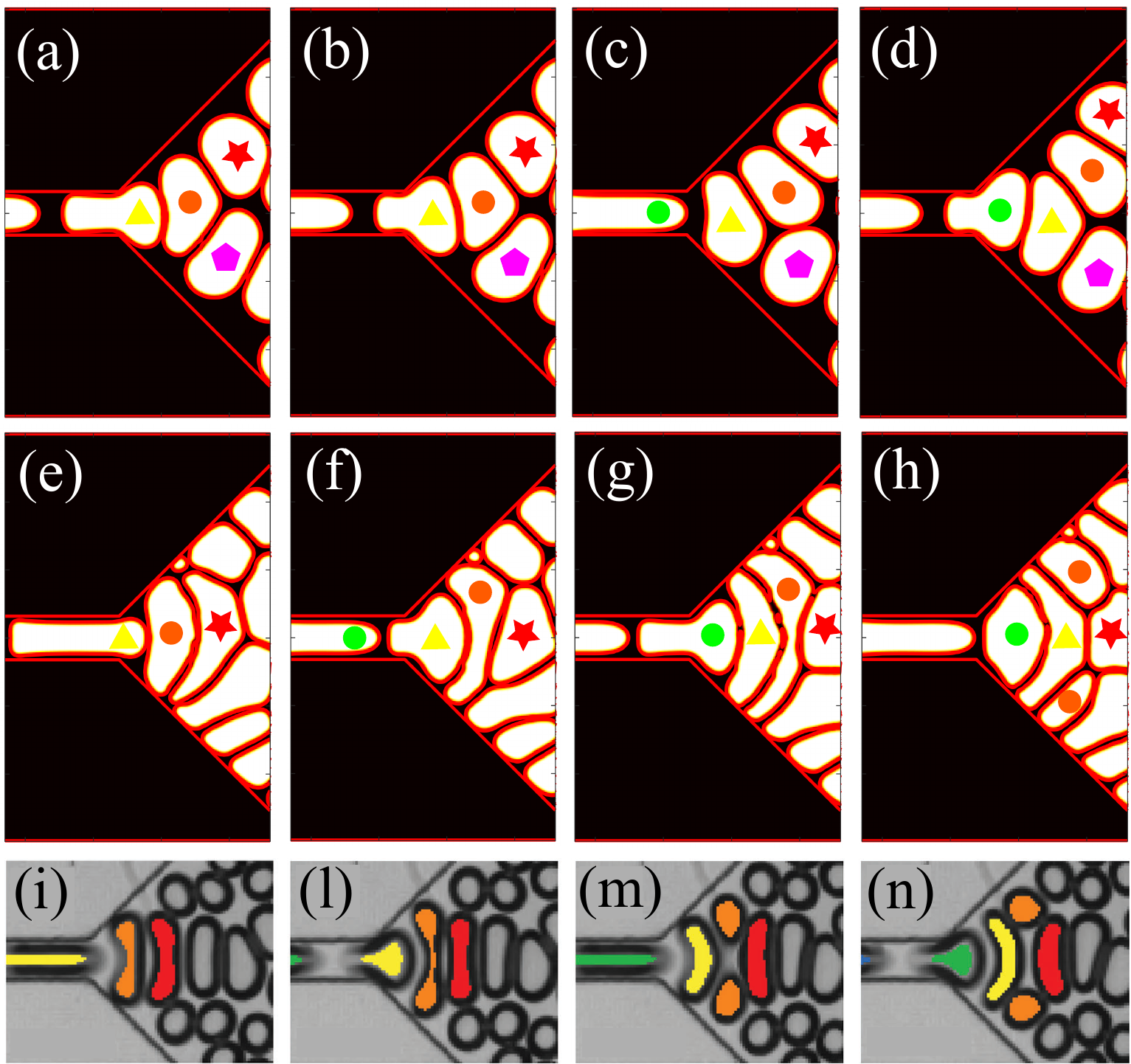}
\caption{(a-d) ($Ca\sim 0.04 $) Push and slide mechanism of the outcoming droplet. 
The yellow-triangle droplet comes out of the channel, pushes the orange-dotted one which, in turn, takes the place of its 
nearest-neighbor droplet (the red-star one). 
(e-h) ($Ca\sim0.16$) Droplet pinch-off process. 
The dotted-orange droplet undergoes a transversal stretching due to the squeezing between the 
outcoming droplet and the red-star drop. 
The stretched droplet finally reaches a critical elongation and thinning under 
the confinement of the neighbor drops before pinch-off.
(i-n) Experimental sequence of the breakup mechanism at $Ca\sim 0.08$ (see \cite{vecchiolla2018bubble}).  
The experimental and numerical critical capillary numbers above which droplets pinch-off can be observed
are $Ca\geq \sim 0.04$ and $Ca \geq \sim  0.05$ respectively. }
\label{fig3}
\end{figure}

As shown in figure \ref{fig3} (a-d), below a given value of the Capillary number at the inlet, 
$Ca \lesssim{0.05}$, every new droplet emerging in the divergent channel pushes 
away another immediately downstream, taking its place in the process. 
Indeed, as clearly sequenced in the figure, the yellow-triangle 
droplet comes out of the channel, pushes the orange-dotted one which, in turn, takes the place of its 
nearest-neighbor droplet (the red-star one). This process is metronomic 
i.e. it does not involve any breakup event and this rythmic {\it push-and-slide} 
mechanism reflects into the regular hexagonal crystal which forms downstream the main channel. 

As stated before, an increase of the Capillary number above a critical value, 
around $Ca\sim0.1$, determines the transition to a heterogeneous, foam-like structure, as shown 
in fig. \ref{fig0}(b).  This latter is due to the subsequent breakup events taking place immediately 
downstream the injection channel, a process  highlighted in figure \ref{fig3} (e-h).

The dotted-orange droplet undergoes a transversal stretching due to the squeezing between the outcoming 
droplet (i.e. the \textit{hammer} droplet ) and the red-star drop (i.e. the \textit{wall} droplet). 
The stretched droplet finally reaches a critical elongation and thinning under the confinement of the neighbor 
drops before pinch-off. In the meantime, the yellow-triangle droplet, due to the rapid slow down determined 
by the channel expansion, gradually takes on a crescent shape, fills the area left free 
by the splitting of the orange drop and becomes a \textit{wall} droplet in turn. 
The splitting mechanism just described is responsible for the formation of smaller droplets, which assemble 
in such a way as to form a heterogeneous foam-like structure within the main channel (fig.\ref{fig0}(b)).

\textcolor{black}{Briefly, what we observe from the simulations is that, frequent and precise pinch-off requires sufficiently high capillary numbers to occur ($Ca > 0.1$ for $\alpha=45^\circ$).
This suggests that the ratio between the viscous forces (extensional force) and surface tension (retraction/restoring force), namely the Capillary number, is likely to govern the behavior of the droplet-droplet pinch-off process. 
Indeed,  as the viscous force retard the expansion of the impinging droplet, the central one stretches and breaks at the midpoint due to the deformation arising from the normal stresses exerted by  the impinging and wall droplet. 
The surface tension then acts so to contrast the effect of the normal stresses, since both the hammer and the wall droplets tend to retract to their undeformed circular state.
}
\textcolor{black}{It is worth noting that a similar pinching mechanism has been recently observed 
experimentally in \cite{vecchiolla2018bubble} in the same range of capillary numbers as in our simulations.Incidentally, the transitional Capillary number of the experiments (i.e. $Ca$ above which the pinching mechanism is observed) was found to be in satisfactory
agreement with the one predicted by the simulations (see caption fig. \ref{fig3}).}

\begin{figure*}
\centering
\includegraphics[scale=0.85]{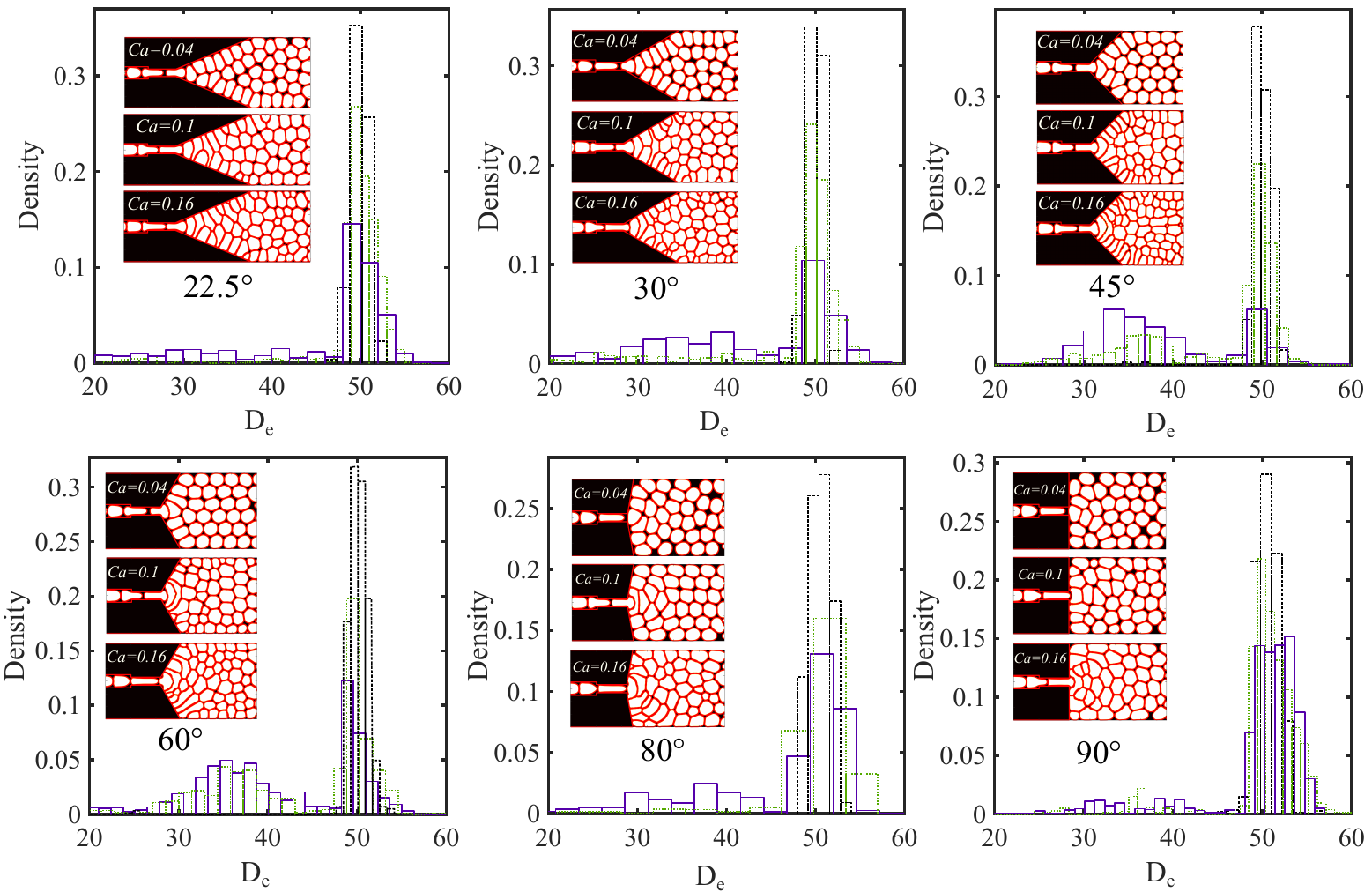}
\caption{Equivalent droplet diameter distributions for each pair of Capillary number and opening angle of the divergent channel ($Ca=0.04$(Dashed line), $Ca=0.1$ (Dotted line) and $Ca=0.16$ (full line)). 
The insets report snapshots of the droplet fields for different values of the Capillary number ($Ca$ increases from top to bottom). \textcolor{black}{The equivalent droplet diameter is the diameter of the circular droplet with the same area of the deformed droplet and can be computed as $D_e=\sqrt{4(A_d/\pi)}$ being $A_d$ the area of the droplet.}}
\label{fig1}
\end{figure*}

At this stage, a question naturally arises as to the role of geometrical details of the 
divergent channel on the wet to dry self-transition.
To address this question, we performed a series of simulations by varying both the Capillary number 
and the opening angle of the divergent channel, so to  systematically assess their combined effect
on the final shape of the assemblies of droplets within the microfluidic channel.

The results of this investigation are summarized in the histograms reported in figure \ref{fig1}. 
Each histogram shows the distribution of the equivalent droplet diameters within the microfluidic 
channel (i.e. the diameter of the circular droplet with the same area of the deformed droplet, \textcolor{black}{computed as $\sqrt{4(A_d/\pi)}$ being $A_d$ the area of the droplet}) 
for a given pair $Ca$ and $\alpha$.

A number of comments is in order :

i) Below $Ca \sim 0.05$, no breakup event is observed, regardless of the opening angle:  the outcoming 
soft structures are monodisperse assemblies of droplets,  as clearly suggested by the dashed-line 
histograms of fig. \ref{fig1}. 

ii) Upon raising the Capillary number, it is possible to trigger the breakup events which lead to the transition between ordered and disordered emulsions. 
By inspecting the histograms ($Ca\sim0.1$ (dotted line) and $Ca\sim0.16$ (solid line) ), it is evident that, for 
a given $\alpha$, the number of breakup events, and in turn, the structure of the resulting emulsion, depend
on the inlet capillary number. Indeed, by increasing the Capillary number, the droplets structure increasingly 
takes the hallmarks of a dense-foam or highly packed dense emulsion (HIPE). 
For $\alpha$ in range $30^\circ -  60^\circ$, $Ca\sim 0.1$ can be regarded as a 
critical value of the Capillary number, around which the emergent soft structure is a 
hybrid between a monodisperse (ordered) and polydisperse (disordered) emulsion, as also evidenced by the the central droplet fields reported in the insets of the histograms.

iii) By further increasing the Capillary number, the assemblies of droplets take a typical foam-like structure, completely 
loosing memory of the structural hexagonal-ordering obtained at lower $Ca$. 
A more complex structure is found, due to the (a) higher degree of deformability of the droplets,  an emergent effect due to the higher values of the Capillary numbers and (b) 
the presence of  smaller droplets which fill the voids between groups of neighbor droplets.

iv) Focusing on the highest value of the Capillary number, $Ca=0.16$, we note that the 
polydispersity, revealed by bimodal histograms, increases as $\alpha$ increases from $22.5^\circ$ to $45^\circ$.
By further increasing the opening angle, the polydispersity starts to recede, nearly 
vanishing at $\alpha=90^\circ$.

\textcolor{black}{
\begin{figure}
\includegraphics[scale=0.7]{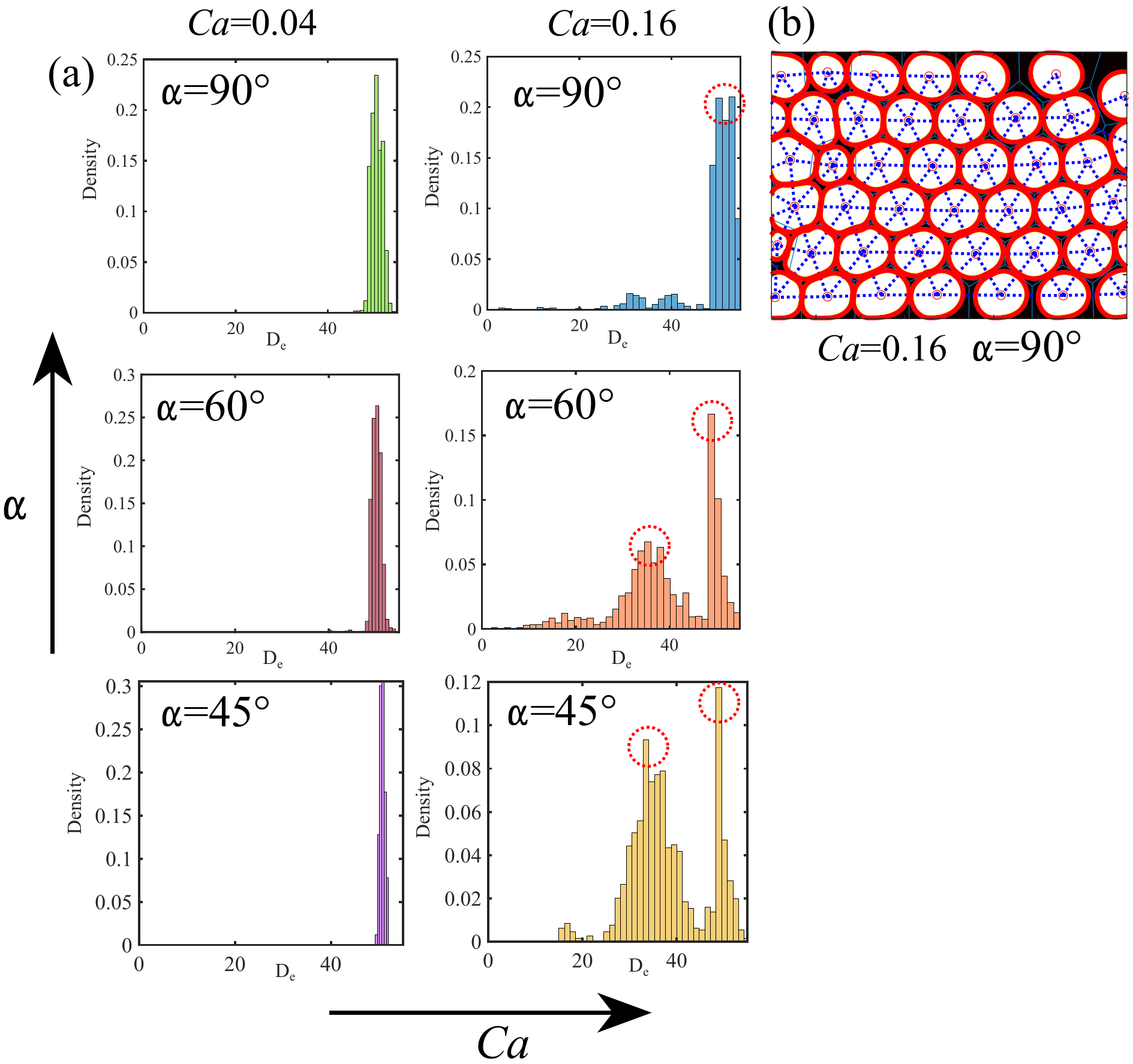}
\caption{(a) Equivalent droplet diameter distributions for two couples of Capillary numbers and opening angles of the divergent channel, namely $Ca=0.04$, $Ca=0.16$ and $\alpha=45^\circ$, $\alpha=60^\circ$, $\alpha=90^\circ$. (b) Droplets' field within the main channel for the case $Ca=0.16$ and $\alpha=90^\circ$  }
\label{sketch}
\end{figure}
To better highlight the aforementioned return towards monodispersity for increasing values of $\alpha$, we directly compare the histograms for six cases, namely, (i)  $Ca=0.04$ and $\alpha=45^\circ$, (ii) $Ca=0.16$ and $\alpha=45^\circ$, (iii)$Ca=0.04$ and $\alpha=60^\circ$,(iv)$Ca=0.16$ and $\alpha=60^\circ$, (v)$Ca=0.04$ and $\alpha=90^\circ$ and (iv)$Ca=0.16$ and $\alpha=90^\circ$ (panel (a)).
A close inspection of the histograms leave no doubt as to the return to monodispersity for the case $Ca=0.16$ and $\alpha=90$.
Indeed, at $Ca=0.16$ and $\alpha=45^\circ$ and $\alpha=60^\circ$  the emulsion is roughly bidisperse as evidenced by the two peaks at $D_e\sim50$ and $D_e\sim 36$ (dashed circles) displayed in the two histograms, the latter one absent in the case $Ca=0.16$ and  $\alpha=90^\circ$.   
The decreasing trend of the ratio between the peaks in the histograms at $Ca=0.16$ ($\sim 1.2$ for $\alpha=45^\circ$ and $\sim 2.5$ for $\alpha=60^\circ$) further points to a gradual return to an ordered structure as the aperture angle increases. This is also apparent from a visual inspection of the droplets' field reported in panel (b) ($Ca=0.16$, $\alpha=90^\circ$) which shows an ensemble of flowing circular droplets of (approximately) the same size.The rare breakup events occurring at the outlet of the injection channel produces a limited number of smaller droplets, an effect evidenced by the small peaks in the upper right histogram.
It is worth noting here that, by a plain argument of mass conservation, polydispersity can arise only as a result of droplet breakup via the droplet hammer mechanism since coalescence is frustrated due to the effect of the near contact forces.
}

\begin{figure}
\centering
\begin{center}
\includegraphics[scale=0.75]{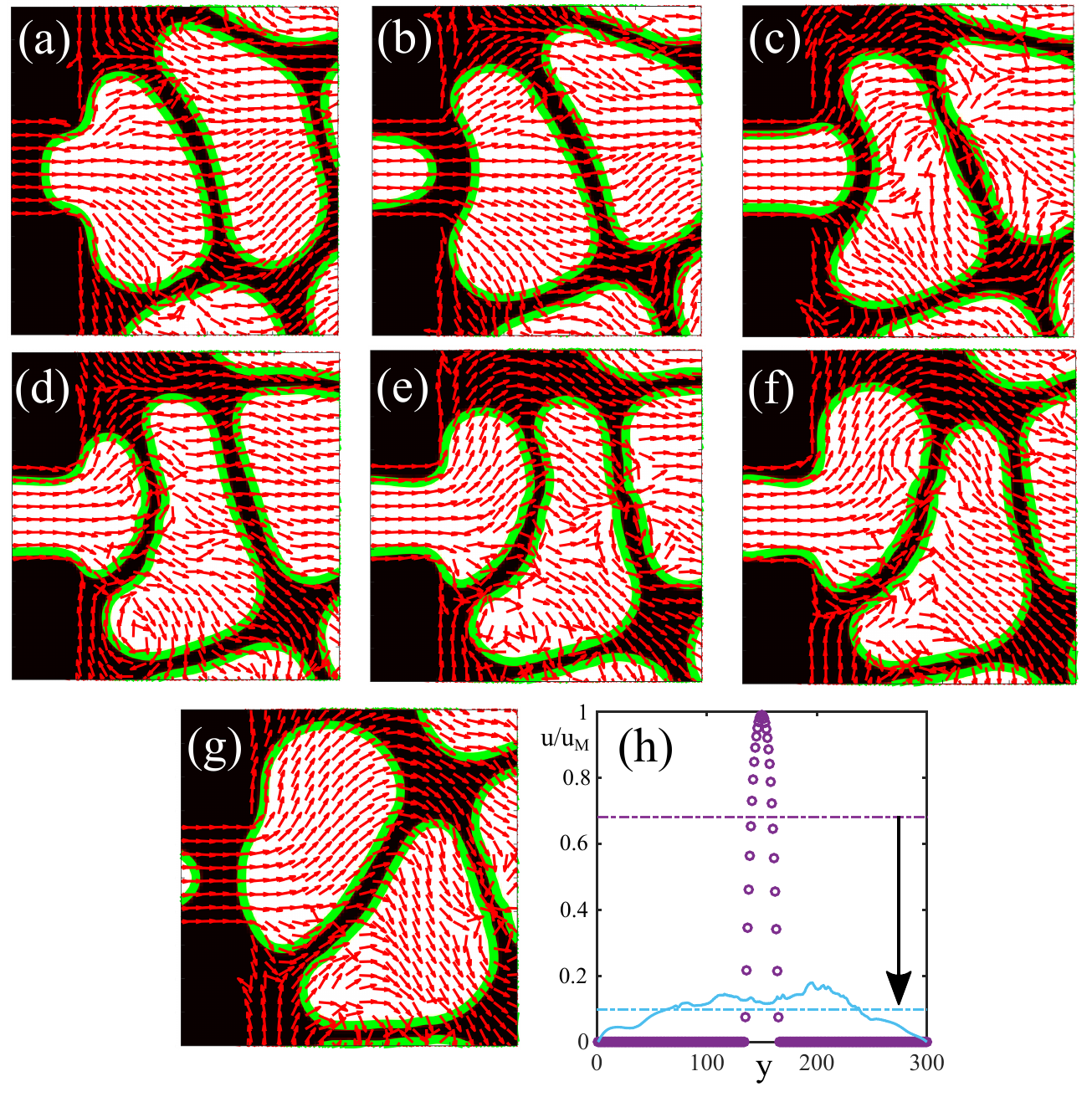}
\caption{(a-g) Droplets' field at the outlet of the injection nozzle for $\alpha=90^\circ$ and $Ca=0.16$ with the normalized vector field superimposed.  
Even at high capillary numbers, the sharp deceleration, clearly evidenced by the velocity profiles reported in panel (h) taken in two distinct sections, within the nozzle(open circles) and at a downstream section,  favors the transversal 
displacement of the outcoming droplets, which preferentially slide on the neighboring droplets rather 
than squeezing them. 
The velocity field is scaled with the maximum flow velocity at the inlet channel. 
The two sections, at which the velocity profiles are evaluated are at $x=190lu$ (inside the injection channel) and $x=340lu$ (within the main channel). The arrow within the plot indicates the average velocity drop between the inlet and the main channel. The axis of the plot are, $u/u_M$ (normalized magnitude of the velocity) versus $y$(crossflow coordinate). }
\label{fig4}
\end{center}
\end{figure}

The counterintuitive behaviour described above, can be intuitively explained as follows:



at high opening angles (approaching to $90 ^\circ$), 
each droplet exiting from the narrow channel experiences a sudden expansion, responsible for a fast recovery 
of their circular shape, just after their emergence within the main channel. 
The fast expansion, in turn, determines a strong deceleration (see plot in fig. \ref{fig4}), which forces the next 
outcoming droplet to loose its \textit{droplet hammer} action, as the opening angle of 
the divergent increases above a critical value between $\alpha=45-60^\circ$.
Further, the sharp deceleration favors the crossflow, transversal displacement of the outcoming droplets, 
which slide preferentially on the downstream, neighbour droplets rather than squeezing them.

The process described above is reported in figure \ref{fig4}.

\textcolor{black}{It is to note that, by varying the surface tension, and by keeping the other parameters fixed (so that the Reynolds number can be kept fixed), it is possible to  vary viscous dissipation over surface forces  independently of the inertia over viscous dissipation ratio.}

Indeed, the viscous \textit{vs} surface tension forces ratio
is responsible for the frequency of breakup events, hence, in turn, for 
the degree of order/disorder observed in the system.
  
This competition strictly depends on the geometrical features of the microfluidic environment.

To be noted that, the importance of Weber and Capillary number over the droplet breakup frequency in microchannels has been also highlighted in a very recent experimental work of Salari et al. \cite{salari2020expansion} reporting power-law scaling of the dropplets' breakup frequencies as a function of the product $WeCa^2$.

To sum up, the simulations suggest that, the dependence of the crystal order on the geometrical feature of the device is 
not one-way since, once the monodispersity and the hexagonal order are lost, they can 
be reclaimed back by either decreasing or increasing the opening angle of the divergent channel 
below/above a critical angle. In other words, either ways, the system looses memory of the disordered configuration.  

\textcolor{black}{As a note, we wish to stress out that, the detection of the specific regime of capillarity in which the transition occurs required an extensive set of simulations on a broad range of Capillary numbers, as such transition was found to occur indeed in a very narrow window of capillarity space. Thus, even though many more $Ca-\alpha$ combinations have been explored, it was found that the cases reported capture the essence of the phenomenon in point.}

To gain a quantitative insight into the order to disorder transition, we introduce a \textit{dispersity} number, $\delta$,  
defined as the ratio between the number of droplets with a diameter below a critical value, $D_{crit}$, and 
the total number of droplets.
This parameter has been evaluated for each pair of Capillary number $Ca$ and opening angle $\alpha$. 

\begin{figure}
\centering
\includegraphics[scale=0.85]{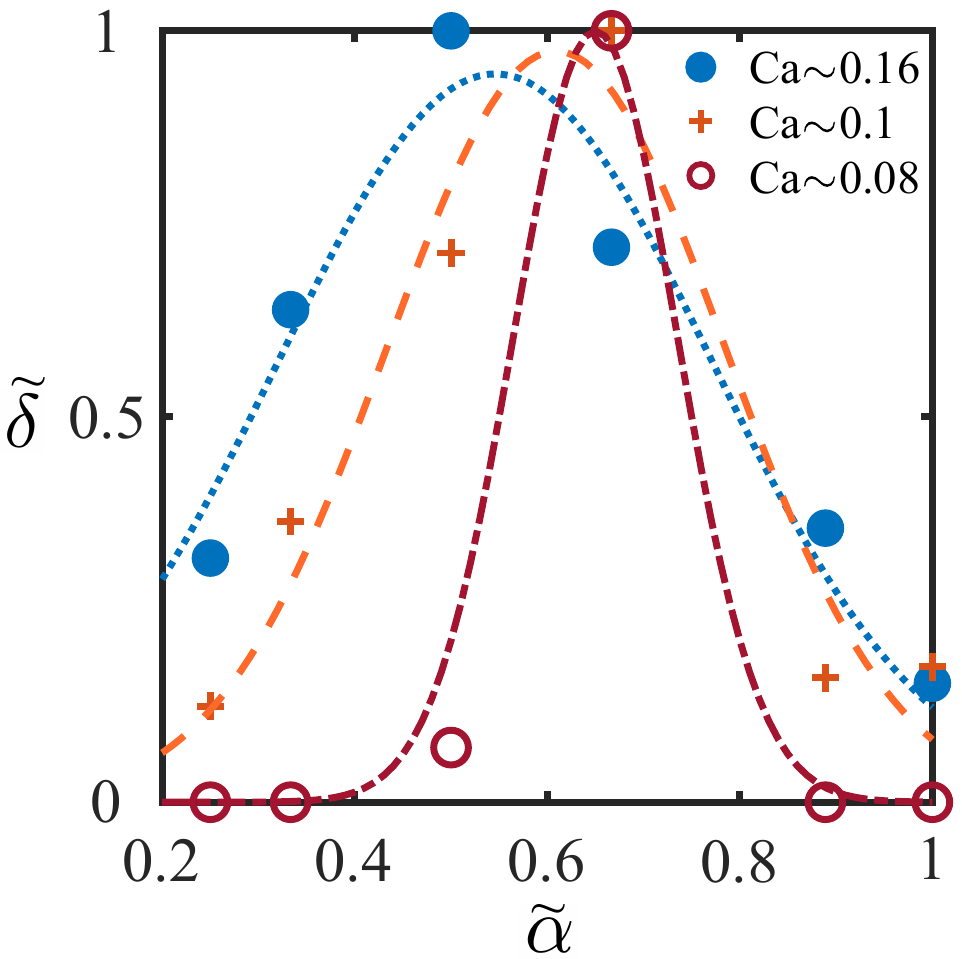}
\caption{ Non dimensional \textit{dispersity} ($\tilde{\delta}$) as a function of the opening angle $\tilde{\alpha}$ for different values of  $Ca$.
Each dispersity set, $\delta(Ca,\tilde{\alpha})$, follows  a Gaussian 
trend, with mean and variance depending on the Capillary number. Fitting function: $\tilde{\delta}(\tilde{\alpha})= e^{-\left((\tilde{\alpha}-\alpha_M)/(2Ca)\right)^2}$}
\label{fig2}
\end{figure}

The plot in figure \ref{fig2} reports these data, made non-dimensional 
by the maximum opening angle $\tilde{\alpha}=\frac{\alpha}{\pi/2}$  
and the maximum value of dispersity  $\tilde{\delta}=\frac{\delta}{\delta_{M}}$, respectively.

A few comments are in order:

The first observation is that each dispersity set, $\delta(Ca,\tilde{\alpha})$, follows  a Gaussian 
trend, with mean and variance depending on the Capillary number:
\begin{equation}
\tilde{\delta}(\tilde{\alpha})=e^{-\left((\tilde{\alpha}-\alpha_M(Ca))/(2Ca)\right)^2}
\end{equation}.
In other words, the dispersity of the system features a \textit{"temperature"} which scales linearly with the inlet Capillary 
number, $T=2Ca$, and a mean value of the opening angle, slightly depending on the Capillary 
number and ranging between $45-60^\circ$. 

The analysis carried out in this paper should be of direct use for experimental research.  
Indeed, for each inlet Capillary number, which can be readily determined by evaluating 
the droplet velocity within the inlet channel, one can single out the channel geometry 
which allows to obtain the desired degree of polydispersity of 
the soft structure, by simply querying the gaussian curves. 

Reciprocally,  given the channel geometry, the capillary number can be tuned in such a way as 
to modify the morphology of the droplet assembly, according again to the gaussian relation provided in this paper.

The present findings are expected to help in defining experimental protocols for 
the development of novel, optimized, low-dimensional, soft porous matrices with tunable properties.
We refer in particular to the so-called functionally graded materials \cite{rabin1995functionally}, namely composite 
materials characterized by a controlled spatial variation of their microstructure, which are capturing mounting
interest for a variety of material science, biology and medical applications.
\section{Conclusions}
In summary, we reported on order to disorder
self-transition in dense emulsions in 
divergent microfluidic channels, as originated by a geometry-controlled competition between viscous dissipation
and interfacial forces. We unveiled a counterintuitive mechanism, namely the 
spontaneous reordering of the emulsion at high Capillary numbers, obtained by increasing 
of the opening angle of the divergent channel. 
Such comeback of order is interpreted as the result
of a subtle balance between viscous dissipation and interfacial forces, straight downstream the inlet channel.
Moreover, We found that the dispersity of the droplet system follows a simple Gaussian law, 
whose temperature is directly proportional to the inlet Capillary number. 
The present findings 
are  expected to offer valuable guidance
for the future development of optimised functional materials with locally tunable properties.
\section*{Acknowledgments}
A. M., M. L., A. T. and S. S. acknowledge funding from the European Research Council under the European Union's Horizon 2020 Framework
Programme (No. FP/2014-2020) ERC Grant Agreement No.739964 (COPMAT).
A.M. acknowledges the ISCRA award SDROMOL (HP10CZXK6R)  under the ISCRA initiative, for 
the availability of high performance computing resources and support.

\section*{Appendix}
\subsection*{Droplets' generation}
\begin{figure}
\includegraphics[scale=0.7]{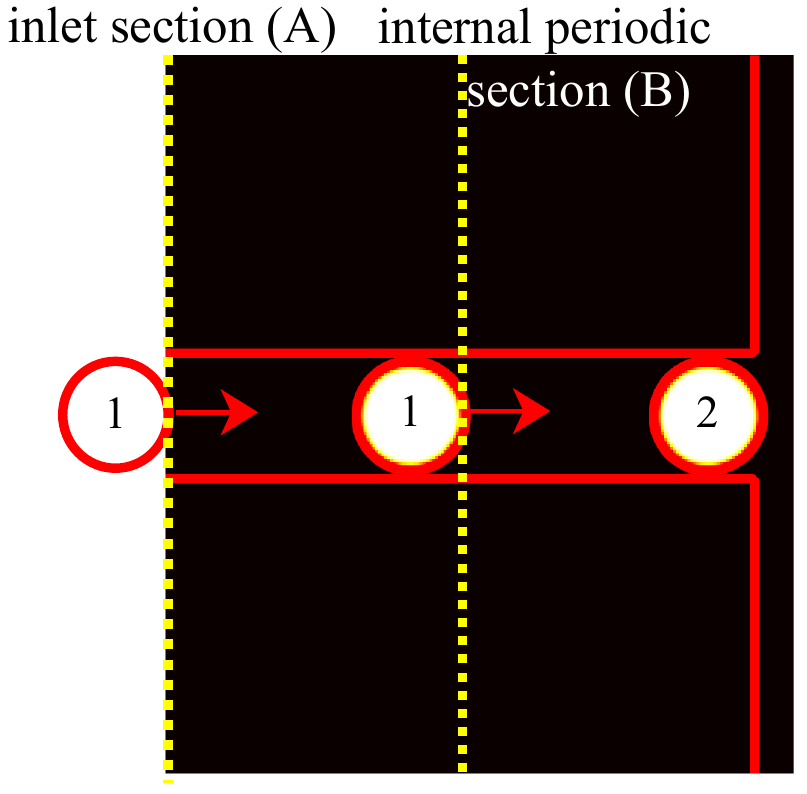}
\caption{Droplet generation via the internal periodic boundary conditions. Populations of  the dispersed and the continuous phase are copied from section (b) back to section (a) and, at the same time, a standard streaming and collision process occurs within the bulk domain. }
\label{sketch_boundary}
\end{figure}

The droplet generation is performed by implementing an \textit{internal} periodic boundary condition which is sketched in figure \ref{sketch_boundary}, for simplicity.
As shown in figure, in order to generate a droplets inflow in the inlet thin channel, the generating region is employed as a source of new droplets. When a droplet passes through section (b), it simultaneosly i) enters into the downstream region  and ii) is copied back to the inlet section by applying periodic boundary conditions from $(b)$ to $(a)$ (see figure \ref{sketch_boundary}).
%

\end{document}